\documentclass[preprint,aps,showpacs]{revtex4-1}
\usepackage{amsmath,amssymb,amsfonts,dcolumn,color,graphicx,graphics,latexsym,placeins,epsfig,tikz}
\usetikzlibrary{arrows,shapes}
\usepackage{subfigure,rotating,hyperref,bm,array}

\newcommand{\be}{\begin{equation}}
\newcommand{\ee}{\end{equation}}
\newcommand{\ba}{\begin{eqnarray}}
\newcommand{\ea}{\end{eqnarray}}

\begin{document}

\title{\Large \bf Wormholes, the weak energy condition, and scalar-tensor
gravity}

\author{Rajibul Shaikh}
\email{rajibulshaikh@cts.iitkgp.ernet.in}
\author{Sayan Kar}
\email{sayan@phy.iitkgp.ernet.in}
\affiliation{${}^{*}$ Centre for Theoretical Studies, Indian Institute of Technology Kharagpur, Kharagpur, 721 302, India.}
\affiliation{${}^{\dagger}$ Department of Physics {\it and} Centre for Theoretical Studies, \\ Indian Institute of Technology Kharagpur, Kharagpur, 721 302, India.}

\begin{abstract}
\noindent We obtain a large class of Lorentzian wormhole spacetimes
in scalar-tensor gravity, for which the matter stress energy does satisfy
the weak energy condition. Our constructions have zero Ricci scalar
and an everywhere finite, non-zero scalar field profile. Interpreting the
scalar-tensor gravity as an effective on-brane theory resulting from
a two-brane Randall--Sundrum model of warped extra dimensions, it is
possible to link wormhole existence with that of extra dimensions.
We study the geometry, matter content, gravitational red-shift and 
circular orbits
in such wormholes and argue that our examples are perhaps among those 
which may have some observational relevance in astrophysics in future. 
We also study traversability and find that our wormholes are indeed
traversable for values of the metric parameters satisfying the 
weak energy condition.
\end{abstract}


\maketitle

\section{Introduction}

\noindent Curved spacetimes for which the Ricci scalar $R$ is identically
zero have been known in General Relativity (GR) since the discovery of the
Schwarzschild solution. While, in GR, the Schwarzschild is a vacuum spacetime for which
both $R$ and $R_{ij}$ are zero, the Reissner-Nordstrom
geometry has $R=0$ but $R_{ij}\neq 0$, thereby implying the presence
of traceless matter. It has been further shown in \cite{R3} that a
generalisation of Schwarzschild spacetime with $\vert g_{00}\vert$
everywhere finite and non-zero and $R=0$ can be obtained. More recently
\cite{R4}, this $R=0$ wormhole spacetime
has been found to be a solution 
in a scalar-tensor theory of gravity which is also the low energy, effective,
on-brane gravity theory for the warped two-brane Randall-Sundrum model.
The $R=0$ spacetime in \cite{R3} when viewed as a solution
in GR, requires matter that violates the Weak Energy Condition (WEC).
However, as shown in \cite{R4}, in the context of the scalar-tensor theory,
the required matter does not violate the WEC and the scalar field
(radion) is also finite and non-zero everywhere. In our work here,
we further generalise the spacetime studied in \cite{R3,R4}.
In particular, newer $R=0$ spacetimes with $\vert g_{00}\vert$ 
everywhere
non-zero are constructed and shown to be solutions 
in scalar-tensor gravity.
A sub-class of such spacetimes are Lorentzian wormholes with the required matter satisfying the WEC.

\noindent In our analysis here, we begin with a general static, spherisymmetric line element
for which the $R=0$ constraint is written as
a differential equation for $\sqrt{\vert g_{00}\vert}=f(r)$. Obviously, 
this differential equation
contains $g_{rr}$ (or, $b(r)$ where $g_{rr}=\frac{1}{1-\frac{b(r)}{r}}$) and its derivatives
in its coefficients. It can therefore be solved once we provide our choice for $b(r)$.
Assuming $b(r)$ as one should for a Lorentzian wormhole, we find that for some specific
choices (eg. the Schwarzschild wormhole and the Ellis--Bronnikov wormhole \cite{EW1,EW2}) 
the differential equation for
$f(r)$ can indeed be solved. We obtain a wide class of geometries in this 
manner. Thereafter, we show how these
geometries fare in terms of the WEC violation issue.  

\noindent As mentioned above, in GR 
we require WEC violation for the matter that threads non-singular wormhole
spacetimes. 
However, it has been shown recently 
that, within GR, a wormhole (which is topologically different from 
the Morris-Thorne class) can be supported by a negative cosmological 
constant (which is not quite exotic, WEC violating matter) \cite{GRW}. 
Further, it is known that 
certain wormhole solutions which have been constructed 
in various modified theories of gravity 
such as Brans-Dicke theory (scalar-tensor gravity with constant coupling $\omega$) \cite{BDW}, $f(R)$-gravity \cite{fRW}, Gauss-Bonnet gravity \cite{GBW}, third-order Lovelock gravity \cite{LGW}, Eddington-inspired Born-Infeld gravity \cite{EiBIW}, mimetic gravity \cite{MW}, DGP gravity \cite{DGPW}, do not require  
WEC violating matter.

\noindent In our work here, we consider scalar-tensor gravity with a 
non-constant coupling $\omega(\Phi)$ and in the presence of matter stress-energy. 
Through our specific solutions, 
we show that there may not be 
WEC violation for the matter required to support a Lorentzian wormhole. 
It must be noted though that
the timelike convergence condition is still violated, as it must be,
for wormholes. The essential point is that the relation between the
convergence condition and the energy condition is not as it is
in GR, because of which there is extra freedom to avoid any violation
in a modified theory of gravity (here, scalar-tensor gravity) \cite{R4}.

\noindent Our paper is organised as follows. In Section \ref{sec:STG}, we briefly recall scalar-tensor gravity both as an independent theory and in the context of braneworld gravity. We provide the details of our construction of $R=0$ spacetimes in Section \ref{sec:solution}. Section \ref{sec:wec} and \ref{sec:m0} deal with the status of the WEC. In Section \ref{sec:redshift}, we study gravitational redshift in the wormhole spacetimes. 
The existence of stable and unstable circular orbits and the 
issue of traversability are addressed in 
Section \ref{sec:traversability}.
Finally, we conclude in Section \ref{sec:conclusion}.

\section{Scalar-tensor gravity}
\label{sec:STG}

\noindent Scalar-tensor theories are well-studied from various angles and 
perspectives. 
The first such theory was, of course, the Jordan-Brans-Dicke theory \cite{jbd}. A scalar field $\Phi$ and
an extra parameter $\omega$ characterises such theories. Though originally considered a constant,
a modified version incorporates a $\Phi$-dependent $\omega$ \cite{ST}. In general, the Einstein field
equations for such theories are given as:
\begin{eqnarray}
G_{\mu\nu}&=& \frac{\kappa}{\Phi} T^{M}_{\mu\nu}+\frac{1}{\Phi}\left(\nabla_{\mu} \nabla_{\nu} \Phi-g_{\mu\nu}\nabla^{\alpha} \nabla_{\alpha} \Phi \right) \nonumber \\
& & +\frac{\omega(\Phi)}{\Phi^2}\left(\nabla_{\mu} \Phi \nabla_{\nu} \Phi-\frac{1}{2}g_{\mu\nu}\nabla^{\alpha} \Phi \nabla_{\alpha} \Phi \right)
\label{eq:field_equation1}
\end{eqnarray}
where the $T^{M}_{\mu\nu}$ is the matter energy-momentum tensor and the metric $g_{\mu\nu}$ is written in the
Jordan frame where the action is not in the standard canonical (Einstein-Hilbert) form. Viewed as an independent
competitor of GR, a scalar tensor theory could be specified by providing a form for $\omega(\Phi)$. 
Note that $\Phi$ must satisfy the equation,  
\begin{equation}
\nabla^{\alpha}\nabla_{\alpha}\Phi=\kappa\frac{T^M}{2\omega(\Phi)+3}-\frac{1}{2\omega(\Phi)+3}\frac{d\omega}{d\Phi}\nabla^{\alpha}\Phi \nabla_{\alpha}\Phi
\label{eq:scalar_equation}
\end{equation}
Various aspects of scalar-tensor theories have been extensively studied and we will not review them here. See \cite{STG} and references therein for detailed reviews. Taking the trace of (\ref{eq:field_equation1}) and using (\ref{eq:scalar_equation}), one obtains
\begin{equation}
R=-\frac{2\kappa\omega T^M}{\Phi(2\omega+3)}-\frac{3}{\Phi(2\omega+3)}\left(\frac{d\omega}{d\Phi}-\frac{\omega(2\omega+3)}{3\Phi}\right)\nabla^\alpha\Phi \nabla_\alpha\Phi
\nonumber
\end{equation}
Note that traceless matter implies $R=0$ in GR. But, in scalar-tensor gravity, traceless matter does not imply $R=0$ in general. But, for the specific form of $\omega(\Phi)$ given by
\begin{equation}
\frac{d\omega}{d\Phi}-\frac{\omega(2\omega+3)}{3\Phi}=0 \hspace{0.2cm} \Rightarrow \hspace{0.2cm} \omega(\Phi)=\frac{3c_0\Phi}{2(1-c_0\Phi)}
\nonumber
\end{equation}
$R=-\kappa c_0 T^M$ and hence, traceless matter implies $R=0$. Here, $c_0$ is an integration constant. As we shall see, the case $c_0=-1$ arises in the low energy, effective on-brane quasi-scalar-tensor gravity theory developed by Kanno and Soda \cite{KS} in the context of 
the Randall-Sundrum two-brane model.

\noindent It is well-known that scalar-tensor theories arise as effective theories of gravity in diverse
contexts. For example, the low-energy effective gravity theory which emerges out of superstring theory
contains a scalar known as the dilaton and the theory is a $\omega=-1$ Brans-Dicke theory \cite{bdstring}. In the braneworld scenario, our four dimensional visible universe is considered as a lower dimensional hypersurface called a brane, which is embedded in a higher dimensional bulk. In the low energy limit, the field equation governing bulk gravity, leads to effective field equations for gravity on the brane. The presence of the extra dimensional bulk leaves its imprint by modifying the Einstein field equations on the brane. The most popular among the on-brane effective Einstein equations was obtained by Shiromizu-Maeda-Sasaki \cite{SMS} in the context of a single-brane model and contains a non-local term (bulk-Weyl dependent $\mathcal{E}_{\mu\nu}$). The on-brane effective Einstein equations obtained by Kanno and Soda \cite{KS} in the context of the two-brane Randall Sundrum model, however, does not contain any non-local contribution. Earlier, the non-local $\mathcal{E_{\mu\nu}}$ was used 
to obtain $R=0$ spacetime solutions \cite{R7,R8,R9,R10} (mostly wormhole solutions) in the single-brane effective 
theory of Shiromizu-Maeda-Sasaki. In the effective on-brane theory of 
Kanno and Soda, we have an effective energy momentum tensor (other than the matter on the branes) constructed from the scalar 
radion field (which measures the inter-brane distance) and its derivatives. The radion field plays a crucial role 
in obtaining both static, spherically symmetric spacetimes and cosmological solutions \cite{R4,R5,R6}. It has been 
shown that the presence of the radion field can correctly reproduce the observed virial mass of galaxy 
clusters and observed galaxy rotation curves. Thus, the radion field can act as a possible dark matter 
candidate \cite{DM}.

\noindent Let us now briefly review the low energy effective on-brane quasi-scalar-tensor theory developed by 
Kanno and Soda \cite{KS}. The action is given as,
\begin{eqnarray}
S &=& \frac{1}{2\kappa^2}\int d^5x\sqrt{-g}\left(\mathcal{R}+\frac{12}{l^2}\right)-\sum\limits_{i=A,B}\sigma_i \int d^4x \sqrt{-g^{i\;brane}} \nonumber \\
& & +\sum\limits_{i=A,B} \int d^4x \sqrt{-g^{i\;brane}}\mathcal{L}^i_{matter}
\nonumber
\end{eqnarray}
where $\kappa^2$, $\mathcal{R}$ and $g_{\mu\nu}^{i\;brane}$ are the five dimensional gravitational coupling constant, the five dimensional curvature scalar and the induced metric on the branes, respectively. $\sigma_A=\frac{6}{\kappa^2 l}$ and $\sigma_B=-\frac{6}{\kappa^2 l}$ are the brane-tensions. The positive tension brane A (Planck brane) and the negative tension brane B (visible brane) are respectively placed at fixed bulk locations ($y=0$ and $y=l$). The bulk line element is,
\begin{equation}
ds^2=e^{2\phi(x^{\mu})}dy^2+g_{\mu\nu}(y,x^\mu)dx^\mu dx^\nu
\nonumber
\end{equation}
where the bulk curvature radius is $l$. Assuming the brane curvature radius $L$ as large compared to the bulk curvature radius $l$, i.e., $\left(\frac{l}{L}\right)^2\ll 1$, Kanno and Soda used a low energy expansion scheme (the gradient expansion method) wherein the bulk metric and the extrinsic curvature are expanded in powers of $\left(\frac{l}{L}\right)^2$. They obtained the effective field equations,
\begin{eqnarray}
G_{\mu\nu}&=&\frac{\kappa^2}{l\Phi}T^{B}_{\mu\nu}+\frac{\kappa^2(1+\Phi)}{l\Phi}T^{A}_{\mu\nu}+\frac{1}{\Phi}\left(\nabla_{\mu} \nabla_{\nu} \Phi-f_{\mu\nu}\nabla^{\alpha} \nabla_{\alpha} \Phi \right) \nonumber \\
& & -\frac{3}{2\Phi(1+\Phi)}\left(\nabla_{\mu} \Phi \nabla_{\nu} \Phi-\frac{1}{2}f_{\mu\nu}\nabla^{\alpha} \Phi \nabla_{\alpha} \Phi \right)
\label{eq:field_equation}
\end{eqnarray}
on the visible brane. Here, $\Phi=e^{\frac{2d}{l}}-1$, $T^{A}_{\mu\nu}$, $T^{B}_{\mu\nu}$ and $f_{\mu\nu}$ are respectively the radion field, the matter on the Planck brane, the matter on the visible brane and the metric tensor on the visible brane. The proper distance $d$ between the branes is defined as
\begin{equation}
d(x)=\int_0^l e^{\phi(x)}dy
\nonumber
\end{equation}
The radion field $\Phi$ satisfies the following equation of motion on the visible brane:
\begin{equation}
\nabla^{\alpha}\nabla_{\alpha}\Phi=\frac{\kappa^2}{l}\frac{(T^A+T^B)}{2\omega+3}-\frac{1}{2\omega+3}\frac{d\omega}{d\Phi}\nabla^{\alpha}\Phi \nabla_{\alpha}\Phi
\end{equation}
where both the traces $T^A$ and $T^B$ are taken with respect to the metric tensor $f_{\mu\nu}$ and the coupling function is given by
\begin{equation}
\omega(\Phi)=-\frac{3\Phi}{2(1+\Phi)}
\label{eq:coupling2}
\end{equation}
Using the radion equation of motion in the trace of the field equation (\ref{eq:field_equation}), we obtain $R=\frac{\kappa^2}{l}T^B$, where $R$ is the curvature scalar on the visible brane. Note that this is the $c_0=-1$ case discussed earlier. Therefore, traceless matter implies $R=0$ in this theory. Using the transformation $\xi=\sqrt{1+\Phi}$, we can express the radion equation of motion in the form
\begin{equation}
\nabla^{\alpha}\nabla_{\alpha}\xi-\frac{R}{6}\xi=\frac{\kappa^2}{6l}T^A\xi
\label{eq:radion_equation1}
\end{equation}

\noindent In the subsequent sections, we obtain our new solutions. It is important to note that the solutions
could either be viewed as solutions in a scalar-tensor theory which has no link with extra dimensions 
or as solutions in the context of braneworld gravity. In obtaining the solutions, we take the Planck brane to be devoid of matter, i.e., $T^A_{\mu\nu}=0$. Further, henceforth, we refer to this on-brane scalar-tensor gravity as Kanno-Soda theory of gravity because of the specific form of the coupling function $\omega(\Phi)$ given in (\ref{eq:coupling2}).

\section{R=0 spacetimes}
\label{sec:solution}

\noindent We begin with the static, spherically symmetric line element given as
\begin{equation}
ds^2=-f^2(r)dt^2+\frac{dr^2}{1-\frac{b(r)}{r}}+r^2\left(d\theta^2+\sin^2\theta d\phi^2\right)
\nonumber
\end{equation}
where $f(r)$ and $b(r)$ are unknown functions to be determined. We are interested in vanishing curvature scalar ($R=0$) solutions. The $R=0$ constraint yields the following second order differential equation for $f(r)$:
\begin{equation}
\left(1-\frac{b}{r}\right)f''(r)+\frac{4r-3b-b'r}{2r^2}f'(r)-\frac{b'}{r^2}f(r)=0
\label{eq:Ricci_flat}
\end{equation}
The above equation can also be re-written as a first order differential equation for $b(r)$. Attempts have been made to 
find out $R=0$ spacetime solutions in the past \cite{R1,R2,R3,R4,R5,R6,R7,R8,R9,R10}. In most of these articles, the authors have (a) considered known forms 
of $f(r)$ and obtained solutions by solving the first order differential 
equation for $b(r)$ or (b) used some known $R=0$ solutions present in the 
literature and analysed them in the context of an effective theory of gravity.  In our work, we take a different route. We specify the form of $b(r)$ and 
solve the second order differential equation for $f(r)$. This way we hope 
to obtain a more general solution different from those obtained earlier. 
Another advantage of specifying the form of the shape function $b(r)$ is 
that we can choose to have wormhole 
solutions with a desired spatial shape. One can take different forms of 
$b(r)$. In our work here, we take $b(r)=2m+\frac{\beta}{r}$. Putting $f(r)=\frac{F(r)}{r}$ in the $R=0$ equation, we obtain
\begin{equation}
\left(1-\frac{2m}{r}-\frac{\beta}{r^2}\right)F''(r)+\left(\frac{m}{r^2}+\frac{\beta}{r^3}\right)F'(r)-\frac{m}{r^3}F(r)=0
\nonumber
\end{equation}
After some manipulations, the above equation can be re-written in the form
\begin{equation}
\frac{d}{dr}\left[\frac{1}{r}(r^2-2m r-\beta)^{3/2}\frac{d}{dr}\left(\frac{F(r)}{\sqrt{r^2-2m r-\beta}}\right)\right]=0
\nonumber
\end{equation}
which can be integrated to obtain
\begin{equation}
f(r)=C_1\left(m+\frac{\beta}{r}\right)+C_2\sqrt{1-\frac{2m}{r}-\frac{\beta}{r^2}}
\end{equation}
where $C_1$ and $C_2$ are integration constants. Therefore, the line element 
becomes
\begin{eqnarray}
ds^2&=&-\left[C_1\left(m +\frac{\beta}{r}\right)+C_2\sqrt{1-\frac{2m}{r}-\frac{\beta}{r^2}} \right]^2dt^2 \nonumber \\
& & +\frac{dr^2}{1-\frac{2m}{r}-\frac{\beta}{r^2}}+r^2d\Omega^2
\label{eq:general_metric}
\end{eqnarray}
It is useful to note that, when $m=0$ or $\beta=0$, the $|g_{00}|$ is of the 
general form, $\left(C_3 b(r)+C_4 \sqrt{1-\frac{b(r)}{r}}\right)^2$. However, 
when $m$, $\beta$ are both non-zero, this generic form does not hold and we 
have $|g_{00}|=\left(C_3 b_1(r)+C_4 \sqrt{1-\frac{b(r)}{r}}\right)^2$. However, it is important to note that, in all cases, the red-shift function is finite 
and non-zero everywhere.

\noindent We now look at special cases of our general solution. Some of these 
special cases reproduce the $R=0$ spacetimes obtained earlier by other authors. For $\beta=0$, $C_1 m=\kappa$ and $C_2=\lambda$, the spacetime takes the form
\begin{equation}
ds^2=-\left[\kappa+\lambda\sqrt{1-\frac{2m}{r}} \right]^2dt^2+\frac{dr^2}{1-\frac{2m}{r}}+r^2d\Omega^2
\end{equation}
This spatially Schwarzschild spacetime has been obtained 
in \cite{R2,R3,R4,R7}. This spacetime represents either a wormhole or a naked singularity for $\kappa\neq 0$. For $\kappa=0$ and $\lambda=1$, the spatially Schwarzschild solution reduces to the Schwarzschild spacetime. Taking $\kappa=1$ and $\lambda=0$ in this solution, we obtain the spatial Schwarzschild wormhole obtained in \cite{R8}. For $2m=\frac{r_0^2}{r_0-M}$, $\beta=-\frac{Mr_0^2}{r_0-M}$, $C_1m=1$ and $C_2=0$, the general metric takes the form
\begin{equation}
ds^2=-\left(1-\frac{2M}{r}\right)^2 dt^2+\frac{dr^2}{\left(1-\frac{r_0}{r}\right)\left(1-\frac{r_1}{r} \right)}+r^2d\Omega^2
\end{equation}
where $r_1=\frac{Mr_0}{r_0-M}$. This metric has been obtained in \cite{R8,R9}. This metric reduces to the extreme Reissner-Nordstr$\ddot{o}$m form for $r_1=r_0$, i.e., $r_0=2M$. The above metric fails to produce spatial Ellis geometry because $r_1=-r_0$ implies $r_0=r_1=0$. However, we can obtain spatial Ellis geometry from our general solution by taking $m=0$ and $\beta=r_0^2$. The line element becomes
\begin{equation}
ds^2=-\left(\frac{C_1 r_0^2}{r}+C_2\sqrt{1-\frac{r_0^2}{r^2}} \right)^2dt^2+\frac{dr^2}{1-\frac{r_0^2}{r^2}}+r^2d\Omega^2
\label{eq:spatial_Ellis}
\end{equation}
This new metric has not been obtained before. Note that, due to the presence of the $\frac{1}{r}$ factor in $|g_{00}|$, the above metric is different from the spatial Ellis geometry obtained in \cite{TG} for vanishing torsion scalar in the context of modified teleparallel gravity. For $\beta=-Q^2$, the general metric reduces to spatial Reissner-Nordstr$\ddot{o}$m metric. The spatial Reissner-Nordstr$\ddot{o}$m metric further reduces to the Reissner-Nordstr$\ddot{o}$m metric for $C_1=0$ and $C_2=1$.

\noindent Our focus is on the general metric (\ref{eq:general_metric}). 
Depending on the signs and values of $m$, $\beta$, $C_1$ and $C_2$, the general spacetime may represent a black hole, a wormhole  or a naked singularity. The nature of the solution depends on the presence of the zeroes of $f(r)$ and $g(r)$.
We will focus on solutions for which $m>0$, $\beta>0$ and $\eta=\frac{C_2}{C_1m}>-1$.
The solution for such choices does represent a Lorentzian wormhole.

\noindent For later use, let us now write down the solution in isotropic
coordinates. The transformation from the Schwarzschild radial coordinate
$r$ to the isotropic coordinate $R$ is given as
\begin{equation}
r=\left (1+\frac{m}{R}+\frac{m^2+\beta}{4R^2}\right)R
\end{equation}
Using the transformation between
$r$ and $R$ we obtain, in isotropic coordinates, the line element,
\begin{equation}
ds^2 = -\frac{h^2(R)}{U^2(R)}dt^2 + U^2(R)\left (dR^2+R^2d\Omega_2^2\right )
\end{equation}
where
\begin{equation}
h(x)= (C_1 m- C_2) \left (q_1+x\right )\left (q_2+x\right )
\end{equation}
\begin{equation}
U(x)=1 + x^2+2\mu x
\end{equation}
and 
\begin{equation}
x=\frac{\sqrt{m^2+\beta}}{2 R} \hspace{0.2in};\hspace{0.2in} \mu = 
\frac{m}{\sqrt{m^2+\beta}}
\end{equation}
The constants $q_1$ and $q_2$ are given as
\begin{equation}
q_1= \frac{\mu(1+\eta)}{1-\sqrt{\mu^2\eta^2+1-\mu^2}}
\hspace{0.2in};\hspace{0.2in}
q_2= \frac{\mu(1+\eta)}{1+\sqrt{\mu^2\eta^2+1-\mu^2}}
\end{equation}
and $\eta = \frac{C_2}{C_1 m}$.
It is easy to show that $\eta$ and $\mu$ are related to $q_1$ and $q_2$ 
as follows:
\begin{equation}
\mu=\frac{q_1 q_2+1}{q_1+q_2} \hspace{0.2in};\hspace{0.2in}
\eta=\frac{q_1 q_2-1}{q_1q_2+1}
\label{eq:mu_eta}
\end{equation}
It can be shown that, for a spacetime without a horizon $\eta>-1$ is necessary. We will use the isotropic coordinate form of the line element
while discussing the weak energy condition.

\noindent Defining $\xi =\sqrt{1+\phi}$ one can solve the radion field equation
as shown in \cite{R4}. Obviously one ends up introducing a new constant
which we denote as $\gamma$. We have
\begin{equation}
\xi = \frac{2\gamma}{q_1-q_2} \log\left|\frac{q_1+x}{q_2+x}\right|
\end{equation}
where we have omitted an overall additive constant. It is easy to see that
$\xi^2-1$ is never zero as along as $q_1>q_2>0$. Note that the domain of
$x$ is from $0$ to $1$, where $x=1$ corresponds to the wormhole throat.   

\noindent We will now turn towards analysing the Weak Energy Condition
inequalities and demonstrate that it can indeed be satisfied for the
wormhole we have constructed.

\section{Weak Energy Condition (WEC)}
\label{sec:wec}
\noindent The Weak Energy Condition comprises the inequalities
\begin{equation}
\rho\geq 0,\hspace{0.3cm} \rho+\tau\geq 0, \hspace{0.3cm} \rho+p\geq 0
\nonumber
\end{equation}
for a diagonal energy momentum tensor with energy density $\rho$, radial pressure $\tau$ and tangential pressures $p$ defined in the static observer's frame. Physically, the WEC means that the matter energy density is always
non-negative in any frame of reference.

\noindent For our case here, we have, for the L.H.S. of the WEC
inequalities,
\begin{eqnarray}
\frac{\kappa^2}{l}\rho = 
\frac{16\mu^2(\mu^2-1) x^4}{m^2 U^2} \left(\xi^2-1\right )
+\frac{16\gamma\mu^2x^4}{m^2(q_1+x)^2(q_2+x)^2}\left [ 
\gamma -  \nonumber \right .\\ \left.  
\xi \,\, \frac{q_1+q_2-2q_1q_2\mu +2 x(1-q_1q_2)+x^2(2\mu-q_1-q_2)}{U}\right ]
\end{eqnarray}

\begin{eqnarray}
\frac{\kappa^2}{l}\left (\rho +\tau \right ) 
=\frac{8\mu^2x^3}{m^2(q_1+x)^2(q_2+x)^2}\left [8\gamma^2 x +4\gamma \xi(q_1q_2-x^2) -
\nonumber \right . \\ \left . \frac{(q_1+q_2+2 x +(q_1+q_2)x^2+2 q_1q_2 x)(q_1
+x)(q_2+x)}{U} (\xi^2-1)\right ]
\end{eqnarray}

\begin{equation}
\frac{\kappa^2}{l}\left (\rho +p \right ) =
\frac{\kappa^2}{l}\left [ 2 \rho - \frac{1}{2} (\rho+\tau)\right]
\end{equation}
where the last equation follows from the traceless requirement on the
matter energy momentum tensor.

\noindent For arbitrary values of the parameters it is not possible to
satisfy these inequalities. However, one can isolate the negativity by
looking at the requirements that emerge near $x=0$. In the limit $x\to 0$, we have
\begin{equation}
\frac{\kappa^2}{l}\rho= \frac{16\mu^2}{m^2}\left[(\mu^2-1)(\xi_0^2-1)+\frac{\gamma^2}{q_1^2 q_2^2}-\frac{\gamma\xi_0}{q_1^2 q_2^2}(q_1+q_2-2\mu q_1 q_2)\right]x^4+\mathcal{O}(x^5)
\end{equation}
\begin{equation}
\frac{\kappa^2}{l}(\rho+\tau)=\frac{8\mu^2}{m^2q_1q_2}\left[-(\xi_0^2-1)(q_1+q_2)+4\gamma\xi_0\right]x^3+\mathcal{O}(x^4)
\end{equation}
\begin{equation}
\frac{\kappa^2}{l}(\rho+p)=-\frac{\kappa^2}{2l}(\rho+\tau)+\mathcal{O}(x^4)
\end{equation}
where $\xi_0=\xi(x=0)$. Note that, at the leading order, (i.e., terms which are $\mathcal{O}(x^3)$), $(\rho+\tau)$ and $(\rho+p)$ are opposite in sign. This violates the WEC in the limit $x\to 0$. Therefore, we must set the coefficient of $x^3$ in $(\rho+\tau)$ to zero. This yields the following expression for $\gamma$.
\begin{equation}
\gamma^2= \frac{(q_1-q_2)^2(q_1+q_2)}{4(q_1+q_2)(\log\frac{q_1}{q_2})^2 -8(\log\frac{q_1}{q_2}) (q_1-q_2)}
\label{eq:gamma1}
\end{equation}
which reduces to the expression quoted in \cite{R4} for $q_1=q$ and $q_2=1$. With the above condition satisfied, it can be shown that in $(\rho+\tau)$, $\mathcal{O}(x^4)$ term vanishes as well and $\mathcal{O}(x^5)$ term is always positive for any positive $q_1$ and $q_2$. Therefore, it is clear from the traceless condition that, at the $\mathcal{O}(x^4)$, positivity of $\rho$ implies positivity of $(\rho+p)$. Note that, at this point, $q_1$ and $q_2$ can be chosen so that $\rho$ and $\gamma^2$ are positive. However, we choose those values of $q_1$ and $q_2$ for which the coefficient of the $\mathcal{O}(x^4)$ term in $\rho$ vanishes. This gives us the following requirement on $\gamma^2$.
\begin{equation}
\gamma^2= \frac{(q_1-q_2)^2 q_1^2 q_2^2 (1-\mu^2)}{4(1-\mu^2)q_1^2q_2^2 (\log\frac{q_1}{q_2})^2 +2 (q_1-q_2) (q_1+q_2-2q_1q_2 \mu) \log\frac{q_1}{q_2} -(q_1-q_2)^2}
\label{eq:gamma2}
\end{equation}
Equating the above two expressions for $\gamma^2$, one arrives at
a relation between $q_1$ and $q_2$ which must be obeyed.
\begin{equation}
\left [ 8q_1^2q_2^2 (1-\mu^2) + 2(q_1+q_2)(q_1+q_2-2q_1q_2\mu)\right ]
\log\frac{q_1}{q_2} = q_1^2-q_2^2
\label{eq:q1q2}
\end{equation}
Thus, once we are able to choose a $q_1$ and a $q_2$ which satisfies the above
relation, we can use their values to find $\gamma$. The set of
values for $q_1$, $q_2$ and $\gamma$ can therefore be used to 
write down the line element and the scalar field profile for a
Lorentzian wormhole in scalar-tensor gravity which will satisfy all the
WEC inequalities.

\noindent The central issue at this point is whether we can obtain an
analytical handle on the relation between $q_1$ and $q_2$. 
To get rid of the logarithm, we can assume $q_1=q_2 e^y$. This will
eventually lead us to a quartic equation in $q_2$ for which we can
obtain the positive, real valued roots in terms of $y$. Choosing $y$ one can then obtain $q_1$, $q_2$ and then $\gamma$. 

\begin{figure}[ht]
\centering
\includegraphics[scale=0.95]{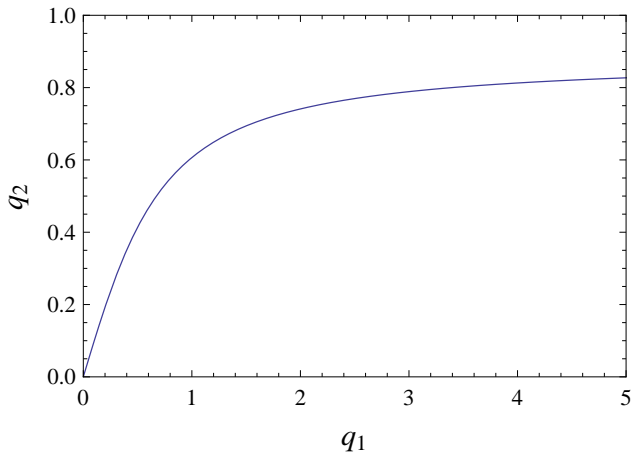}
\caption{Plot of $q_2$ as a function of $q_1$.}
\label{fig:q}
\end{figure}

\begin{figure}[ht]
\centering
\includegraphics[scale=0.95]{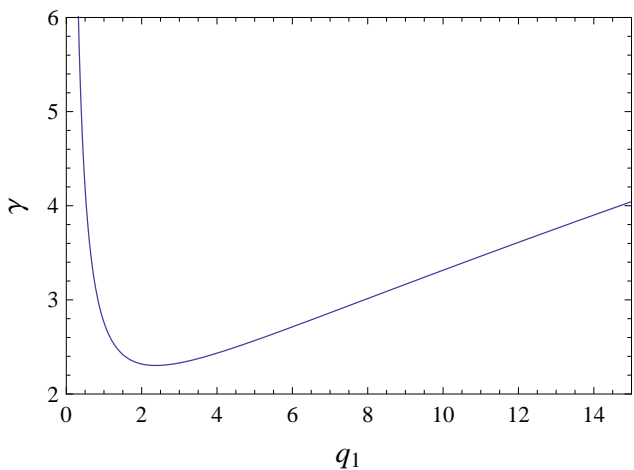}
\caption{Plot showing $\gamma$ as a function of $q_1$.}
\label{fig:gamma}
\end{figure}

\begin{figure}[ht]
\centering
\includegraphics[scale=0.95]{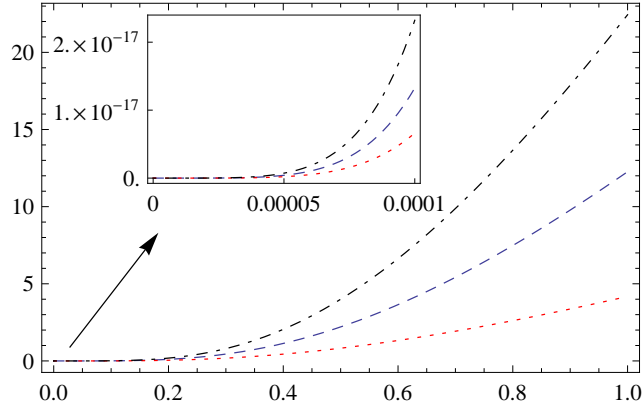}
\caption{Plots of $\frac{\kappa^2}{l}\rho$ (blue dashed curve), $\frac{\kappa^2}{l}(\rho+\tau)$ (red dotted curve) and $\frac{\kappa^2}{l}(\rho+p)$ (black dot-dashed curve) as a function of $x$ for $y=0.2948$ which corresponds to $\eta\simeq -0.5$, $q_1\simeq 0.669$, $q_2\simeq 0.498$, $\gamma\simeq 3.419$ and $\mu\simeq 1.142$.}
\label{fig:EC1}
\end{figure}

\begin{figure}[ht]
\centering
\includegraphics[scale=0.95]{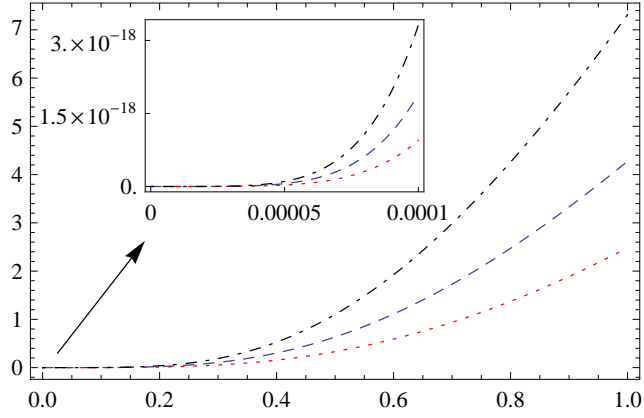}
\caption{Plots of $\frac{\kappa^2}{l}\rho$ (blue dashed curve), $\frac{\kappa^2}{l}(\rho+\tau)$ (red dotted curve) and $\frac{\kappa^2}{l}(\rho+p)$ (black dot-dashed curve) as a function of $x$ for $y=0.7475$ which corresponds to $\eta\simeq 0.0$, $q_1\simeq 1.453$, $q_2\simeq 0.688$, $\gamma\simeq 2.437$ and $\mu\simeq 0.934$.}
\label{fig:EC2}
\end{figure}

\begin{figure}[ht]
\centering
\includegraphics[scale=0.95]{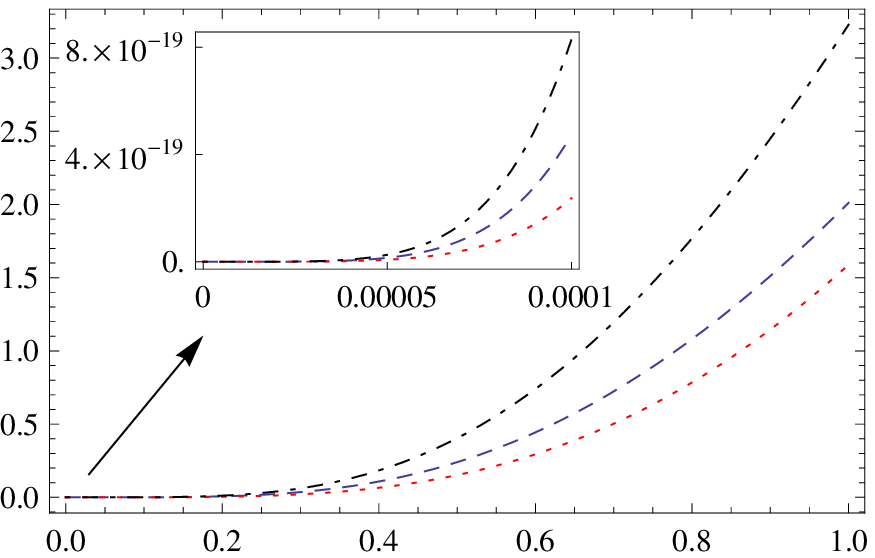}
\caption{Plots of $\frac{\kappa^2}{l}\rho$ (blue dashed curve), $\frac{\kappa^2}{l}(\rho+\tau)$ (red dotted curve) and $\frac{\kappa^2}{l}(\rho+p)$ (black dot-dashed curve) as a function of $x$ for $y=1.5267$ which corresponds to $\eta\simeq 0.5$, $q_1\simeq 3.716$, $q_2\simeq 0.807$, $\gamma\simeq 2.4$ and $\mu\simeq 0.884$.}
\label{fig:EC3}
\end{figure}

\noindent We will now illustrate the above statements with examples and also
demonstrate our claim that the WEC is indeed satisfied for the wormholes we have
constructed. Putting $q_1=q_2 e^y$ in (\ref{eq:q1q2}), we obtain
\begin{equation}
A q_2^4+B q_2^2+C=0 \hspace{0.2cm}\Rightarrow \hspace{0.2cm} q_2^2=\frac{-B\pm \sqrt{B^2-4AC}}{2A}
\label{eq:q2}
\end{equation}
where,
\begin{equation}
A=-4, \hspace{0.5cm} B=2\left(e^{-y}-1\right)^2,
\end{equation}
\begin{equation}
C=\left(e^{-2y}-1\right)^2+2e^{-y}\left(e^{-2y}+1\right)-\frac{1}{2y}\left(e^{-y}+1\right)^2\left(1-e^{-2y}\right)
\end{equation}
Therefore, we fix $q_1$ and $q_2$ and hence $\mu$, $\eta$ and $\gamma$ by choosing a particular value of $y$. It can be shown that $-1<\eta< 1$ for the whole range $-\infty<y<\infty$. Therefore, the definitions of $q_1$ and $q_2$ indicate that both are positive. Figure \ref{fig:q} shows the parametric plot for $q_1$ and $q_2$. Figure \ref{fig:gamma} shows the dependence of $\gamma$ on $q_1$. Figures \ref{fig:EC1}-\ref{fig:EC3} show the L.H.S of the WEC inequalities for three sets of parameter values. It is to be noted that WEC inequalities are satisfied. 



\section{The $m=0$ limit and WEC}
\label{sec:m0}
\begin{figure}[ht]
\centering
\includegraphics[scale=0.95]{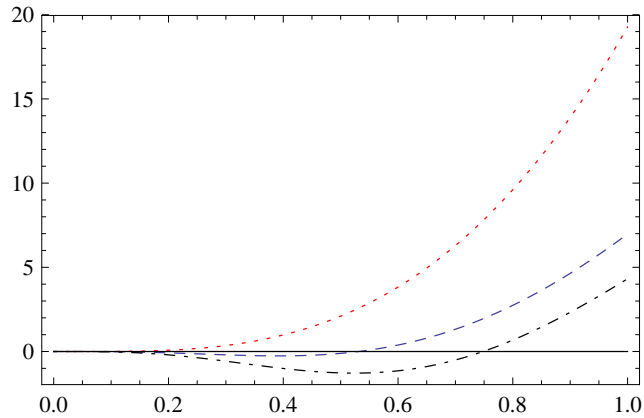}
\caption{Plots of $\frac{\kappa^2}{l}\rho$ (blue dashed curve), $\frac{\kappa^2}{l}(\rho+\tau)$ (red dotted curve) and $\frac{\kappa^2}{l}(\rho+p)$ (black dot-dashed curve) as a function of $x$ for $\delta=0.5$.}
\label{fig:EC4}
\end{figure}

\noindent Let us now take the limit $m\to 0$. In this limit $\mu\to 0$ and $\eta$ diverges. But, the product $\delta=\mu\eta$ is finite. In terms of $\delta$, we have
\begin{equation}
q_1=\frac{\delta}{1-\sqrt{1+\delta^2}}, \hspace{0.3cm} q_2=\frac{\delta}{1+\sqrt{1+\delta^2}}, \hspace{0.3cm} q_1 q_2=-1
\end{equation}
Therefore, $q_1$ and $q_2$ have opposite signs. In this limit, the factor $\frac{\mu^2}{m^2}$ in the energy density and pressures must be replaced by $\frac{1}{\beta}$. It can be shown that $h(x)$ does not have zeroes in the range $0\leq x\leq 1$ for $\delta>0$, thereby representing a wormhole. But, for $\delta<0$, it does have a zero and represents a naked singularity. Let us first consider the case $\delta>0$. Like the $m\neq 0$ case, here also, one must consider the expression (\ref{eq:gamma1}). With this, it can be shown that, in the limit $x\to 0$, the coefficient of the $\mathcal{O}(x^4)$ term in $\rho$ cannot be set to zero since it is negative for all $\delta>0$, thereby violating the WEC. It also implies that the expression in (\ref{eq:gamma2}) is not valid. However, we can make use of the freedom to add an additive constant to $\xi$ and make this coefficient positive. If we do so, then we find that $\gamma^2<0$ in (\ref{eq:gamma1}), i.e., $\gamma$ is imaginary. On the other hand, it can be shown that if we force $\gamma^2$ to be positive, then either $\rho$ or $(\rho+\tau)$ or both become negative in the limit $x\to 0$. Therefore, the wormholes without the mass term `$m$' violate the WEC. Figure \ref{fig:EC4} shows the WEC violation for $m=0$. We have carried out a similar analysis for the naked singularity case $(\delta<0)$ where we found that the WEC can be satisfied in the limit $x\to 0$, but the radion field $\xi$ vanishes at a point between the singularity and spatial infinity. This gives an invalid solution. Therefore, WEC cannot be satisfied for the naked singularity.

\section{Gravitational redshift}
\label{sec:redshift}

\begin{figure}[ht]
\centering
\includegraphics[scale=0.95]{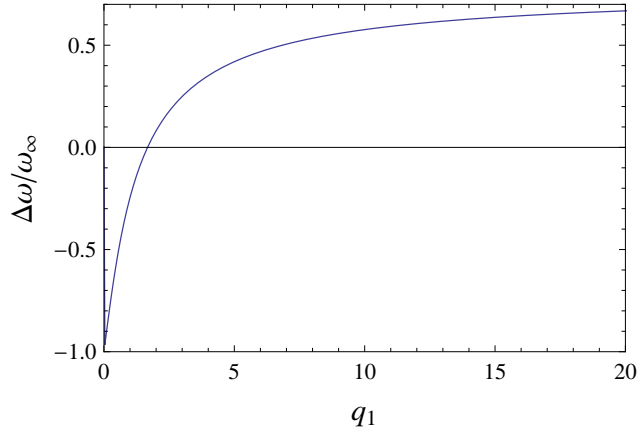}
\caption{Plot showing the fractional change in frequency suffered by a signal emitted from the wormhole throat and propagating to spatial infinity, as a function of $q_1$ with $q_2$ and $q_1$ satisfying Eqn. (29).}
\label{fig:redshift}
\end{figure}

\noindent Let us now look at the gravitational redshift of a 
light signal propagating in the $R=0$ wormhole spacetimes discussed
above. Note that by virtue of having a $g_{00}$ which is everywhere finite
and non-zero, the gravitational redshift is always finite. 
By definition, wormhole spacetimes are horizon-free. 
Assume that a light signal is emitted at a frequency $\omega_{0}$ 
from the wormhole throat ($x=1$) and it travels to spatial infinity ($x=0$) 
where a static observer receives it at a frequency $\omega_\infty$. 
Therefore, the fractional change in frequency due to the gravitational 
redshift is given by the standard formula.
\begin{equation}
\frac{\Delta\omega}{\omega_\infty}=\frac{|g_{00}(x=0)|^{1/2}}{|g_{00}(x=1)|^{1/2}}-1=\frac{2q_1q_2(1+\mu)}{(q_1+1)(q_2+1)}-1
\end{equation}
We choose  values of $q_1$ and $q_2$ which satisfy Eqn. (\ref{eq:q1q2}) 
and hence, the WEC. Figure \ref{fig:redshift} shows the fractional change 
in frequency as a function of $q_1$. It is worth noting that the light signal 
may have a negative gravitational redshift, i.e., a  
gravitational blueshift ($\Delta \omega<0$) for certain values of $q_1$ and 
$q_2$. For the three sets of $q_1$ and $q_2$ given in Figs. \ref{fig:EC1}-\ref{fig:EC3}, $\frac{\Delta\omega}{\omega_\infty}$ are respectively $-0.429$, $-0.066$ and $0.3265$. For $q_1\simeq 1.672$ and $q_2\simeq 0.713$, $\frac{\Delta\omega}{\omega_\infty}$ vanishes. Therefore, the wormholes can exhibit both 
redshift and 
blueshift. It can be shown that, for those $q_1$ and $q_2$ satisfying the WEC, 
the maximum redshift is $\frac{\Delta\omega}{\omega_\infty}\big\vert_{max}=\left(\sqrt{1+\sqrt{5}}-1\right)\simeq 0.798$. This redshift does not 
exceed $80\%$ for such values of $q_1$ and $q_2$. 
However, we can have more than $80\%$ redshift if we choose $q_1$ and $q_2$ arbitrarily, which obviously violates the WEC. 
Thus, in a certain sense, as mentioned above, 
the WEC does restrict the gravitational redshift to an amount below $80\%$.
It is possible to use this result (along with other signatures) 
to identify whether the cause of an observed gravitational frequency shift
is indeed due to a wormhole.

\section{Effective potentials, circular orbits and traversability}
\label{sec:traversability}

\subsection{Effective potentials and circular orbits}

\begin{figure}[ht]
\centering
\subfigure[$l=1.0$]{\includegraphics[scale=0.90]{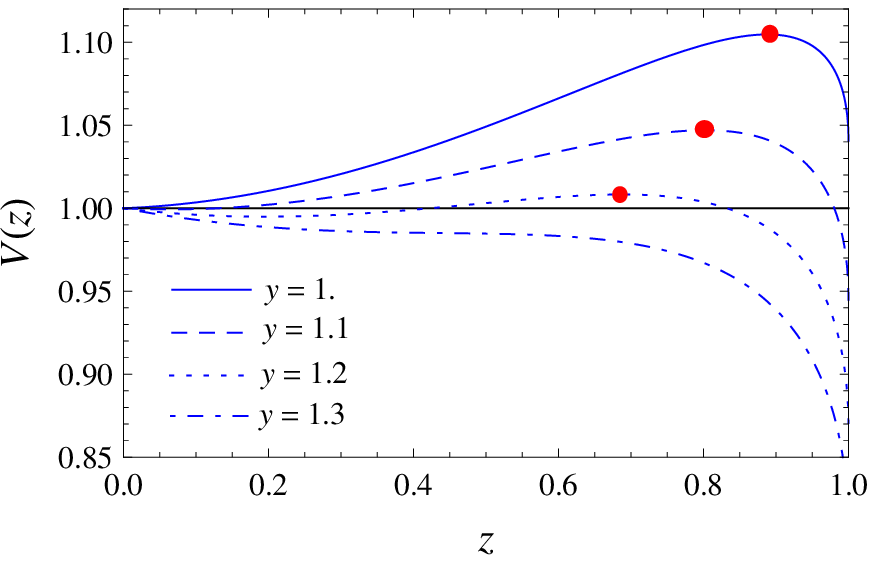}}
\subfigure[$l=1.0$]{\includegraphics[scale=0.90]{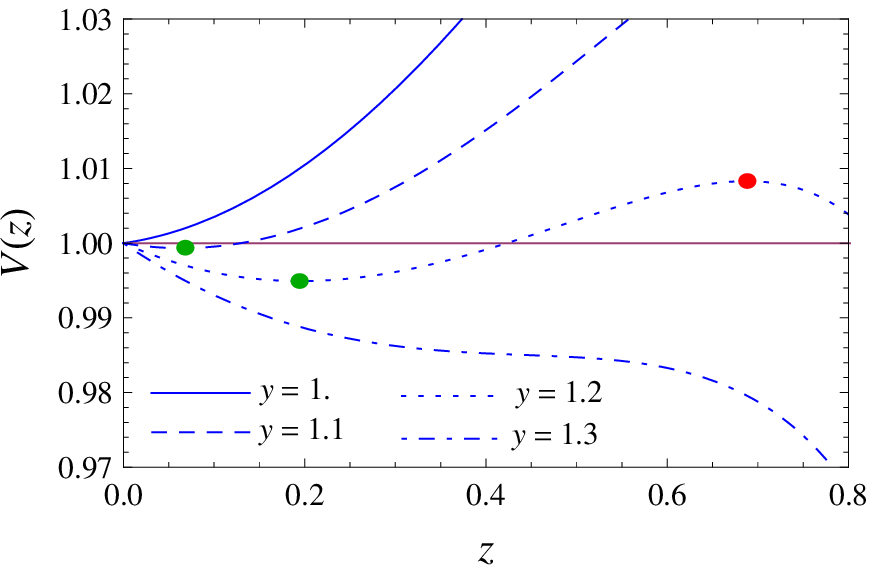}}
\subfigure[$y=1.2$]{\includegraphics[scale=0.90]{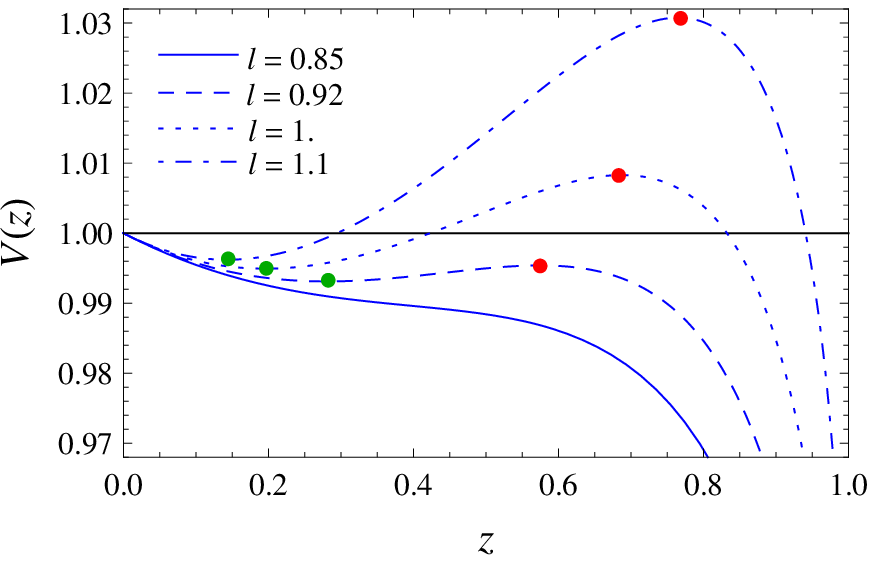}}
\subfigure[$y=1.2$]{\includegraphics[scale=0.90]{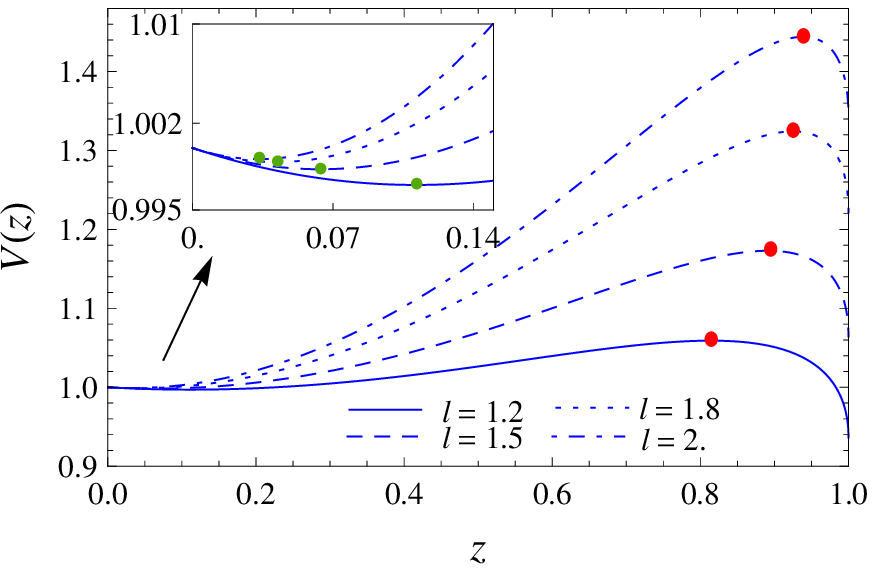}}
\caption{Plots of the effective potential for timelike geodesics for (a)$\&$(b) different $y$ and (c)$\&$(d) different $l$. (b) is the zoomed in version of (a). The green and red dots represent respectively the stable and unstable circular orbits.}
\label{fig:potential1}
\end{figure}

\begin{figure}[ht]
\centering
\includegraphics[scale=0.95]{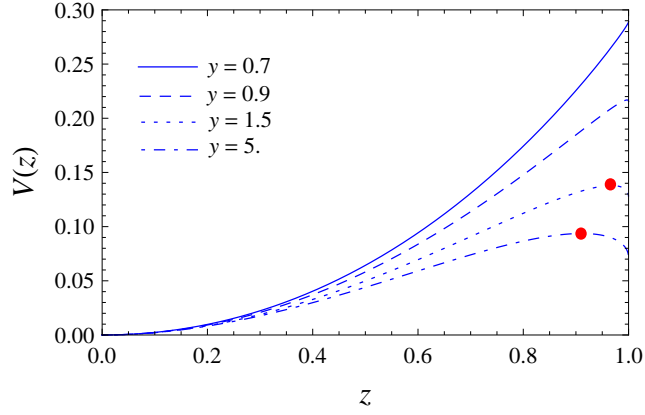}
\caption{Plots of the effective potential for null geodesics for different $y$. Red dots represent unstable circular orbits. Here, we have taken $l=1.0$.}
\label{fig:potential2}
\end{figure}

\begin{figure}[ht]
\centering
\includegraphics[scale=0.95]{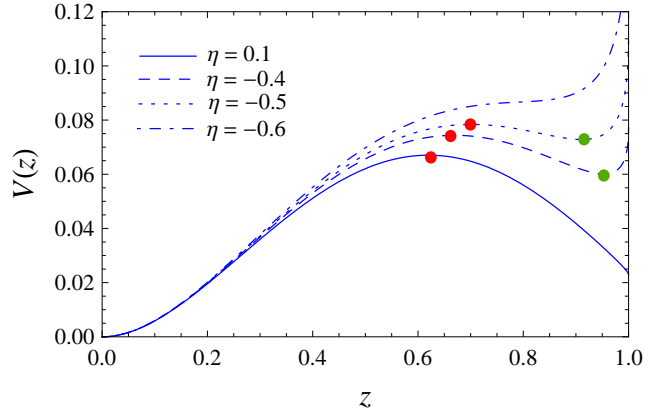}
\caption{Plots of the effective potential for null geodesics for different $\eta$. The green and red dots represent respectively the stable and unstable circular orbits. Here, we have taken $\mu=5.0$ and $l=1.0$.}
\label{fig:potential3}
\end{figure}

\noindent The geodesic equations for a test particle moving in the 
equatorial ($\theta=\frac{\pi}{2}$) plane of the spacetime \ref{eq:general_metric} reduce to the following first integrals:
\begin{equation}
\dot{t}=\frac{E}{f^2(r)}, \hspace{0.3cm} \dot{\phi}=\frac{L}{r^2}, \hspace{0.3cm} \frac{f^2(r)\dot{r}^2}{1-\frac{b(r)}{r}}+V(r)=E^2
\end{equation}
where the energy $E$ and the angular momentum $L$ are constants of motion. 
An overdot represents differentiation with respect to the parameter
$\lambda$ (affine for timelike and arbitrary for null geodesics). 
The effective potential $V(r)$ is given by
\begin{equation}
V(r)=\left(-s+\frac{L^2}{r^2}\right)f^2(r)
\end{equation}
where the normalisation constant $s=\dot{x}_\mu \dot{x}^\mu$ is $-1$ 
for timelike geodesics and is $0$ for null geodesics. We now look for 
circular orbits which correspond to $\dot{r}=0$ (i.e. $V=E^2$) and $V'(r)=0$ (i.e. extrema of the effective potential). Minima ($V''(r)>0$) and maxima ($V''(r)<0$) of the effective potential correspond to the stable and unstable circular orbits, respectively. For the line element in (\ref{eq:general_metric}), the 
effective potential can be written as
\begin{equation}
V(z)=\left[-s+\frac{\mu^2 l^2}{(1+\mu)^2}z^2\right]f^2(z)
\end{equation}
where
\begin{equation}
f(z)=\frac{1}{(1+\eta)}\left(1+\frac{1-\mu}{\mu}z+\eta\sqrt{1-\frac{2\mu}{1+\mu}z-\frac{1-\mu}{1+\mu}z^2} \right),
\end{equation}
$l=\frac{L}{m}$ and $z=\frac{r_0}{r}=\frac{m+\sqrt{m^2+\beta}}{r}=\frac{m(1+\mu)}{\mu r}$, $r_0$ being the throat radius. Here, we have normalised $f(r)$ by dividing it by $(C_1\alpha+C_2)$ such that $|g_{tt}|\to 1$ as $r\to \infty$. 
The effective potential is plotted in Figs. \ref{fig:potential1} and \ref{fig:potential2} for the parameter values satisfying the WEC. By choosing $y$, we find
$q_2$ from (\ref{eq:q2}), $q_1$ from $q_1=q_2 e^y$ and hence $\mu$ and $\eta$ from (\ref{eq:mu_eta}). It should be noted that both stable and unstable circular orbits exist for the timelike case. However, for the null case, no stable circular orbit exists for the parameter values satisfying the WEC. But, 
stable circular orbits exist if we choose the parameter values arbitrarily 
which, of course, violates the WEC. To check this, we first note that the 
extremum $z_{ex}$ of the potential are given by
\begin{equation}
\frac{1}{f(z_{ex})}\frac{df}{dz}\Big\vert_{z_{ex}}+\frac{1}{z_{ex}}=0
\label{eq:extremum}
\end{equation}
Using Eqs. (\ref{eq:Ricci_flat}) and (\ref{eq:extremum}), it can be shown that
\begin{equation}
\frac{d^2V}{dz^2}\Big\vert_{z_{ex}}=-\frac{4l^2\mu^3}{(1+\mu)^3}\frac{f^2(z_{ex})}{1-\frac{2\mu}{1+\mu}z_{ex}-\frac{1-\mu}{1+\mu}z_{ex}^2}\left(\frac{1+\mu}{\mu}-\frac{3}{2}z_{ex}\right)
\end{equation}
Therefore, minima ($\frac{d^2V}{dz^2}>0$) exist if $z_{ex}$ are real and $z_{ex}>\frac{2(1+\mu)}{3\mu}$. Note that we must have $0\leq z_{ex}\leq 1$. This is possible when $\mu>2$. Figure \ref{fig:potential3} shows the plots for the 
potential for parameter values chosen arbitrarily. We have checked that 
the WEC is violated for these parameter values. Therefore, stable null circular orbits exist if the WEC is violated.
It should be noted that, for the parameter values used in Fig. \ref{fig:potential3}, $q_1$ and $q_2$ become complex conjugates of each other because the 
square root terms in the expressions of $q_1$ and $q_2$ become imaginary. 
But, the metric function $h(x)$ still remains real since $(q_1+q_2)$ and $q_1 q_2$ are real. It can also be shown that $\xi(x)$ is real. To verify this, 
let $q_1=a+ib$ where $a$ and $b$ are real. Therefore, $q_2=a-ib$. Writing $(q_1+x)=\sqrt{(a+x)^2+b^2}e^{i\theta(x)}$ and $(q_2+x)=\sqrt{(a+x)^2+b^2}e^{-i\theta(x)}$ where $\tan\theta(x)=\frac{b}{a+x}$, we obtain
\begin{equation}
\xi(x)=\frac{2\gamma}{2ib}\log\left[\frac{a+ib+x}{a-ib+x}\right]=\frac{2\gamma\theta(x)}{b}
\end{equation}
which is real.

\subsection{Traversability}
\noindent Traversability of a wormhole demands that the tidal force felt by 
a human traveller moving radially, must be within tolerable limits. 
In an orthonormal basis $\{e_{\hat{0}'}, e_{\hat{1}'}, e_{\hat{2}'}, e_{\hat{3}'}\}$ attached to the traveller frame, the tidal acceleration between 
two parts of his/her body, separated by the deviation vector 
$\xi^{\hat{i}'}$ is given by \cite{morris1}
\begin{equation}
\Delta a^{\hat{j}'}=-c^2 R^{\hat{j}'}_{\; \hat{0}' \hat{k}' \hat{0}'} \xi^{\hat{k}'},
\end{equation}
where $R^{\hat{i}'}_{\; \hat{j}' \hat{k}' \hat{l}'}$ is the Riemann tensor. 
At the throat, the components of the tidal acceleration are given by
\begin{equation}
\Delta a^{\hat{1}'}\big|_{r_0} = \left\{
  \begin{array}{lr}
    \frac{\beta c^2}{r_0^4}\xi^{\hat{1}'} & : C_1\neq 0\\
 \left(\frac{2m}{r_0^3}+\frac{3\beta}{r_0^4}\right)c^2 \xi^{\hat{1}'} & : C_1=0
  \end{array}
\right. .
\label{eq:constraint2}
\end{equation}
\begin{equation}
\Delta a^{\hat{2}',\hat{3}'}\big|_{r_0}=\frac{\sqrt{m^2+\beta}}{r_0^3}\bar{\gamma}_0^2 v_0^2\xi^{\hat{2}',\hat{3}'}
\end{equation}
where $\bar{\gamma}=\left(1-\frac{v^2}{c^2}\right)^{-\frac{1}{2}}$ and $v=\pm\frac{\sqrt{g_{rr}}dr}{\sqrt{|g_{tt}|}dt}$ is the the radial velocity of the 
traveler as measured by a static observer. $v_0$ denotes the radial velocity at the throat. For the wormhole solutions, we must have $C_1\neq 0$. Note that, 
the radial component of the tidal acceleration vanishes for $\beta=0$. Therefore, it can be made arbitrarily small by choosing an appropriate value of $\beta$. We now restrict the radial component below one Earth gravity, i.e., $|\Delta a^{\hat{1}'}|\leq g$.  Thus, for a traveller of typical size $|\xi|\sim$ $2$ 
meters, we obtain
\begin{equation}
m\geq \sqrt{\frac{2c^2}{g}}\sqrt{\frac{\mu^2|(1-\mu)|}{(1+\mu)^3}}
\label{eq:m_constraint}
\end{equation}
\begin{equation}
\beta \left\{
  \begin{array}{lr}
    \geq \frac{2c^2}{g}\frac{|(1-\mu)|(1-\mu)}{(1+\mu)^2} & : \mu\leq 1\\
    \leq \frac{2c^2}{g}\frac{|(1-\mu)|(1-\mu)}{(1+\mu)^2} & : \mu\geq 1
  \end{array}
\right. .
\end{equation}
where we have used the expressions $\mu=\frac{m}{\sqrt{m^2+\beta}}$ and $r_0=m+\sqrt{m^2+\beta}=\frac{m(1+\mu)}{\mu}$. Note that $m$ and $\beta$ are related through $\mu$. For a given metric parameter $\mu$, the constraint on $m$ (and hence on $\beta$) puts a constraint on the throat radius. 
The constraint on $r_0$ is given by
\begin{equation}
r_0\geq \sqrt{\frac{2c^2}{g}}\sqrt{\frac{|(1-\mu)|}{(1+\mu)}}
\end{equation}
For $\mu=0.934$ which corresponds to Fig. \ref{fig:EC2}, we obtain $r_0\geq 3.9R_{E}$, where $R_E=6400$ km is the Earth radius. However, we can reduce the lower limit on $r_0$ if we choose $\mu$ to be close to $1$.

\noindent If the wormhole is to be traversable, not only the tidal acceleration but also the magnitude of acceleration felt by the traveller as he or she travels through it, must be within tolerable limits. In the orthonormal basis of the traveler frame, the magnitude of the acceleration is given by \cite{morris1}
\begin{equation}
a=\mp \frac{1}{f(r)}\sqrt{1-\frac{b(r)}{r}}(\bar{\gamma} f(r))'c^2
\end{equation}
Let us now consider both $\bar{\gamma}$ and $\bar{\gamma}'$ to be 
finite at the throat. Therefore, at the throat $r_0$, we have
\begin{equation}
\big|a\big|_{r_0}=\frac{|C_2|}{|C_1|r_0^2}\bar{\gamma}_0 c^2=\frac{m|\eta|}{r_0^2}\bar{\gamma}_0 c^2
\end{equation}
For the wormhole solutions, we must have $\eta>-1$. Therefore, for a given $r_0$ and $m$ which restrict the tidal acceleration below one Earth gravity, the magnitude of the acceleration can be made arbitrarily small by taking $\eta$ to be very close to zero which, of course, satisfies the WEC (Fig. \ref{fig:EC2}).

\section{Conclusion}
\label{sec:conclusion}

\noindent In this work, we have obtained a class of static, spherically symmetric $R=0$ spacetimes
which generalise certain known $R=0$ line elements. We have shown that our spacetimes can  
arise in a scalar-tensor theory of gravity, an example of which is the
low energy, effective on-brane gravity theory developed by Kanno and Soda \cite{KS} in the context of 
the Randall-Sundrum two-brane model. Some special cases of the general solution reproduce the $R=0$ spacetimes 
obtained earlier by other authors in the context of GR and in the on-brane 
effective gravity theory 
due to Shiromizu-Maeda-Sasaki \cite{SMS} for a 
single-brane scenario. A sub-class of the general spacetimes 
found by us
represents $R=0$ Lorentzian wormholes. We have studied the 
WEC for the matter that supports such wormholes and have shown that they can 
satisfy WEC. This is in contrast to GR, where such wormholes must violate WEC. 
We have also shown that the mass term in the metric functions is necessary 
to satisfy the WEC for these wormholes. Note that most of the wormhole solutions obtained in \cite{BDW} are in vacuum-Brans-Dicke theory. However, the solutions we have obtained include both matter and a non-constant, i.e., a $\Phi$-dependent coupling $\omega(\Phi)$. As a additional curiosity, we also looked at the
WEC for the naked singularity with the parameter $m=0$, and have 
found that it is violated. 

\noindent Apart from constructing solutions and checking the WEC, we 
have calculated the gravitational redshift of a light signal traveling in
such wormholes and have found that the WEC 
restricts the amount of redshift to be below $80\%$. 
Furthermore, we obtained the effective potentials for particle motion and
investigated  the possibility of circular orbits in these wormhole spacetimes. 
It has been shown that both stable and unstable circular orbits exist for the 
timelike case. For the null case, only unstable circular orbits are 
possible if the WEC is satisfied. But, stable null circular orbits exist if the WEC is violated. Finally, we analysed 
traversability in such wormholes and found that they are traversable 
for values of the metric parameters satisfying the WEC. Therefore,
we now have wormholes which satisfy the WEC and are traversable. 
In addition, they may also be viewed as on-brane spacetimes if one 
considers the existence of warped extra dimensions of the Randall-Sundrum
variety, in a two-brane context. In such a context the scalar field (radion)
is related to the inter-brane distance and we have seen that this
can always be kept finite and non-zero everywhere.

\noindent Recently, it has been argued that the universal ringdown 
waveforms from a binary 
coalescence indicate the presence of light rings, rather than of 
horizons 
\cite{ringdown1}. Therefore, an object (e.g. a traversable wormhole) with a 
light ring will display a similar initial ringdown stage (waveform), even 
when its quasinormal mode spectrum is completely different from that of a 
black hole. However, their late time ringdown stages are different \cite{ringdown1,ringdown2}. It has further been shown that, if one does not use the 
thin-shell wormhole (the one used in \cite{ringdown1}) obtained by 
special matching near the Schwarzschild radius but prefers a wormhole 
configuration constructed without 
thin-shells instead, a wormhole may ring like a black hole or 
differently at all times, depending on the values of the wormhole and 
black hole parameters \cite{ringdown3}. This leads to the hope that the 
Lorentzian wormholes we have obtained here, may be relevant in such 
studies and may have an observational consequence
in astrophysics, in future. 

\noindent Further, it is to be noted that the $R=0$ spacetime solution we 
have obtained can be used in the context of any other theory of gravity 
(different from the one considered here) to study the relation between the 
energy conditions and wormholes. We have noted that a sub-class of our $R=0$ 
spacetimes represents naked singularities. Therefore, as a observational test, 
it will be interesting to study gravitational lensing in the spacetimes
we have obtained here and to see whether we can find ways to 
distinguish between 
wormholes and naked singularities \cite{lensing} through observations.

\section*{Acknowledgement}
\noindent R. S. acknowledges the Council of Scientific and Industrial Research, India, for providing support through a fellowship.


\begin{thebibliography}{99}
\bibitem{R3} N. Dadhich, S. Kar, S. Mukherjee, and M. Visser, {\em $R=0$ spacetimes and self-dual Lorentzian wormholes}, Phys. Rev. D {\bf 65}, 064004 (2002).
\bibitem{R4} S. Kar, S. Lahiri, and S. SenGupta, {\em Can extra dimensional effects allow wormholes without exotic matter?}, Phys. Lett. B {\bf 750}, 319 (2015).
\bibitem{EW1}  H. G. Ellis, {\em Ether flow through a drainhole: A particle model in general relativity}, J. Math.  Phys. {\bf 14}, 104 (1973); Errata: J. Math.  Phys. {\bf 15}, 520 (1974).
\bibitem{EW2} K. A. Bronnikov, {\em Scalar-tensor theory and scalar charge}, Acta Phys. Pol. B {\bf 4}, 251 (1973).
\bibitem{GRW} E. Ayon-Beato, F. Canfora, and J. Zanelli, {\em Analytic self-gravitating Skyrmions, cosmological bounces and wormholes}, Phys. Lett. B {\bf 752}, 201 (2016).
\bibitem{BDW} C. H. Brans, {\em Mach's Principle and a relativistic theory of gravitation. II}, Phys. Rev. {\bf 125}, 2194 (1962);
 \newline
 A. G. Agnese and M. LaCamera, {\em Wormholes in the Brans-Dicke theory of gravitation}, Phys. Rev. D {\bf 51}, 2011 (1995);
 \newline
 K. K. Nandi, A. Islam, and J. Evans, {\em Brans wormholes}, Phys. Rev. D {\bf 55}, 2497 (1997);
 \newline
 K. K. Nandi, B. Bhattacharjee, S. M. K. Alam and J. Evans, {\em Brans-Dicke wormholes in the Jordan and Einstein frames}, Phys. Rev. D {\bf 57}, 823 (1998);
 \newline
 E. F. Eiroa, M. G. Richarte, and C. Simeone, {\em Thin-shell wormholes in Brans–Dicke gravity}, Phys. Lett. A {\bf 373}, 1 (2008);
 \newline
 E. F. Eiroa, M. G. Richarte, and C. Simeone, {\em Erratum to ``Thin-shell wormholes in Brans–Dicke gravity" [Phys. Lett. A 373, 1 (2008)]}, Phys. Lett. A {\bf 373}, 2399 (2009);
 \newline
 A. Bhattacharya, I. Nigmatzyanov, R. Izmailov and K. K. Nandi, {\em Brans-Dicke wormhole revisited}, Classical Quantum Gravity {\bf 26} 235017 (2009);
 \newline
 F. S. N. Lobo and M. A. Oliveira, {\em General class of vacuum Brans-Dicke wormholes}, Phys. Rev. D {\bf 81}, 067501 (2010);
 \newline
 S. V. Sushkov and S. M. Kozyrev, {\em Composite vacuum Brans-Dicke wormholes}, Phys. Rev. D {\bf 84}, 124026 (2011);
 \newline
 X. Yue and S. Gao, {\em Stability of Brans-Dicke thin-shell wormholes}, Phys. Lett. A {\bf 375}, 2193 (2011);
\bibitem{fRW} F. S. N. Lobo and M. A. Oliveira, {\em Wormhole geometries in $f(R)$ modified theories of gravity}, Phys. Rev. D {\bf 80}, 104012 (2009);
 \newline
 T. Harko, F. S. N. Lobo, M. K. Mak, and S. V. Sushkov, {\em Modified-gravity wormholes without exotic matter}, Phys. Rev. D {\bf 87}, 067504 (2013);
 
\bibitem{GBW} B. Bhawal and S. Kar, {\em Lorentzian wormholes in Einstein-Gauss-Bonnet theory}, Phys. Rev. D {\bf 46}, 2464 (1992);
 \newline
 M. G. Richarte and C. Simeone, {\em Thin-shell wormholes supported by ordinary matter in Einstein-Gauss-Bonnet gravity}, Phys. Rev. D {\bf 76}, 087502 (2007);
 \newline
 M. G. Richarte and C. Simeone, {\em Erratum: Thin-shell wormholes supported by ordinary matter in Einstein-Gauss-Bonnet gravity [Phys. Rev. D 76, 087502 (2007)]}, Phys. Rev. D {\bf 77}, 089903(E) (2008);
 \newline
 H. Maeda and M. Nozawa, {\em Static and symmetric wormholes respecting energy conditions in Einstein-Gauss-Bonnet gravity}, Phys. Rev. D {\bf 78}, 024005 (2008);
 \newline
 J. Ponce de Leon, {\em Static wormholes on the brane inspired by Kaluza-Klein gravity}, J. Cosmol. Astropart. Phys. {\bf 11} (2009) 013;
 \newline
 M. H. Dehghani and S. H. Hendi, {\em Wormhole solutions in Gauss-Bonnet-Born-Infeld gravity}, Gen. Relativ. Gravit. {\bf 41}, 1853 (2009);
 \newline
 P. Kanti, B. Kleihaus, J. Kunz, {\em Wormholes in dilatonic Einstein-Gauss-Bonnet theory}, Phys .Rev. Lett. {\bf 107}, 271101 (2011);
 \newline
 P. Kanti, B. Kleihaus, J. Kunz, {\em Stable Lorentzian wormholes in dilatonic Einstein-Gauss-Bonnet theory}, Phys. Rev. D {\bf 85}, 044007 (2012);
 \newline
 M. R. Mehdizadeh, M. K. Zangeneh and F. S. N. Lobo, {\em Einstein-Gauss-Bonnet traversable wormholes satisfying the weak energy condition}, Phys. Rev. D {\bf 91}, 084004 (2015);
 \newline
 T. Kokubu, H. Maeda, and T. Harada, {\em Does the Gauss-Bonnet term stabilize wormholes?}, Classical Quantum Gravity {\bf 32} 235021 (2015).
\bibitem{LGW} M. K. Zangeneh, F. S. N. Lobo, and M. H. Dehghani, {\em Traversable wormholes satisfying the weak energy condition in third-order Lovelock gravity},  Phys. Rev. D {\bf 92}, 124049 (2015);
 \newline
 M. R. Mehdizadeh and F. S. N. Lobo, {\em Novel third-order Lovelock wormhole solutions}, Phys. Rev. D {\bf 93}, 124014 (2016).
\bibitem{EiBIW} R. Shaikh, {\em Lorentzian wormholes in Eddington-inspired Born-Infeld gravity}, Phys. Rev. D {\bf 92}, 024015 (2015).
\bibitem{MW} R. Myrzakulov, L. Sebastiani, S. Vagnozzi, and S. Zerbini, {\em Static spherically symmetric solutions in mimetic gravity: rotation curves $\&$ wormholes}, Classical Quantum Gravity {\bf 33}, 125005 (2016).
\bibitem{DGPW} M. G. Richarte, {\em Wormholes and solitonic shells in five-dimensional DGP theory}, Phys. Rev. D {\bf 82}, 044021 (2010);
 \newline
 M. G. Richarte, {\em Cylindrical wormholes in DGP gravity}, Phys. Rev. D {\bf 87}, 067503 (2013);
 \newline
 Y. Tomikawa, T. Shiromizu, and K. Izumi, {\em Wormhole on DGP brane},  Phys. Rev. D {\bf 90}, 126001 (2014).
\bibitem{jbd} C. Brans and R. H. Dicke, {\em Mach's principle and a relativistic theory of gravitation}, Phys. Rev. {\bf 124}, 925 (1961).
\bibitem{ST} P. G. Bergmann, {\em Comments on the scalar-tensor theory}, Int. J. Theor. Phys.{\bf  1}, 25 (1968).
\bibitem{STG} Y. Fujii, K. Maeda, {\em The Scalar-Tensor Theory of Gravitation} (Cambridge University Press, Cambridge, England, 2003);
 \newline 
 V. Faraoni, {\em Cosmology in Scalar-Tensor Gravity} (Springer, New York, 2004);
 \newline
 T. P. Sotiriou, {\em Gravity and scalar fields}, Lect. Notes Phys. {\bf 892}, 3 (2015);
 \newline
 T. P. Sotiriou, {\em Black holes and scalar fields} , Classical Quantum Gravity {\bf 32}, 214002 (2015).
\bibitem{KS} S. Kanno and J. Soda, {\em Radion and holographic brane gravity}, Phys.Rev. D {\bf 66}, 083506 (2002).
\bibitem{bdstring} G. T. Horowitz, {\em The dark side of string theory: Black holes and black strings}, arXiv:hep-th/9210119.
\bibitem{SMS} T. Shiromizu, K. Maeda and M. Sasaki, {\em The Einstein equations on the 3-brane world}, Phys. Rev. D {\bf 62}, 024012 (2000).
\bibitem{R7} R. Casadio, A. Fabbri, and L. Mazzacurati, {\em New black holes in the brane world?}, Phys. Rev. D {\bf 65}, 084040 (2001).
\bibitem{R8} K. A. Bronnikov and S.-W. Kim, {\em Possible wormholes in a brane world}, Phys. Rev. D {\bf 67}, 064027 (2003).
\bibitem{R9} K. A. Bronnikov, V. N. Melnikov and Heinz Dehnen, {\em General class of brane-world black holes}, Phys. Rev. D {\bf 68}, 024025 (2003).
\bibitem{R10} F. Parsaei and N. Riazi, {\em New wormhole solutions on the brane}, Phys. Rev. D {\bf 91}, 024015 (2015).
\bibitem{R5} S. Kar, S. Lahiri, and S. SenGupta, {\em A note on spherically symmetric, static spacetimes in Kanno-Soda on-brane gravity}, Gen. Relativ. Gravit. {\bf 47}, 70 (2015).
\bibitem{R6} S. Kar, S. Lahiri, and S. SenGupta, {\em Radion stability and induced, on-brane geometries in an effective scalar-tensor theory of gravity}, Phys. Rev. D {\bf 88}, 123509 (2013).
\bibitem{DM} S. Chakraborty and S. SenGupta, {\em Radion as a possible dark matter candidate}, arXiv:1511.00646.
\bibitem{R1} S. Kar and D. Sahdev, {\em Restricted class of traversable wormholes with traceless matter}, Phys. Rev. D {\bf 52}, 2030 (1995).
\bibitem{R2} N. Dadhich, {\em Spherically symmetric empty space and its dual in general relativity}, Curr. Sci. {\bf 78}, 1118 (2000) (arXiv:gr-qc/0003018).
\bibitem{TG} C. G. B$\ddot{o}$hmer, T. Harko, and F. S. N. Lobo, {\em Wormhole geometries in modified teleparallel gravity and the energy conditions}, Phys. Rev. D {\bf 85}, 044033 (2012).
\bibitem{morris1} M. S. Morris and K. S. Thorne, {\em Wormholes in spacetime and their use for interstellar travel: A tool for teaching general relativity}, Am. J. Phys. {\bf 56}, 395 (1988).
\bibitem{ringdown1} V. Cardoso, E. Franzin, and P. Pani, {\em Is the gravitational-wave ringdown a probe of the event horizon?}, Phys. Rev. Lett. {\bf 116}, 171101 (2016).
\bibitem{ringdown2} R. A. Konoplya and C. Molina, {\em Ringing wormholes}, Phys. Rev. D {\bf 71}, 124009 (2005).
\bibitem{ringdown3} R. A. Konoplya and  A. Zhidenko, {\em Wormholes versus black holes: quasinormal ringing at early and late times}, arXiv:1606.00517.
\bibitem{lensing} R. Shaikh and S. Kar, in preparation.

\end{thebibliography}
\end{document}